\documentclass[epj]{svjour}

\usepackage{graphicx}
\usepackage{fancyhdr}
\usepackage{amsmath,amssymb}
\def\makeheadbox{{
 }}
\def\mytitle{My title} 
\def\myauthors{My name}  
\def\mytype{My type of session}
\def\mysession{My session}

\def\mytitle{Counting BPS Solitons and Applications} 
\def\myauthors{Kazutoshi Ohta}    
\def\mytype{Parallel Talk}    
\def\mysession{Theoretical Models}

\newcommand{\be}{\begin{equation}}
\newcommand{\ee}{\end{equation}}
\newcommand{\bea}{\begin{eqnarray}}
\newcommand{\eea}{\end{eqnarray}}
\newcommand{\beann}{\begin{eqnarray*}}
\newcommand{\eeann}{\end{eqnarray*}}

\newcommand{\ba}{\begin{array}}
\newcommand{\ea}{\end{array}}

\newcommand{\Z}{\mathbb{Z}}
\newcommand{\R}{\mathbb{R}}
\newcommand{\C}{\mathbb{C}}

\newcommand{\NC}{N_c}
\newcommand{\NF}{N_f}

\begin{document}
\title{Counting BPS Solitons and Applications}
\author{Kazutoshi Ohta\inst{1}%
\thanks{\emph{Email:} kohta\_at\_tuhep.phys.tohoku.ac.jp}%
}                     
%
%
\institute{Department of Physics, 
Graduate School of Science,
Tohoku University,
Sendai 980-8578, Japan
}
%
\date{}
\abstract{
We propose a novel and simple method of computing
the volume of 
the moduli space of BPS solitons in supersymmetric gauge theory. We use a D-brane realization of vortices and 
T-duality relation to domain walls. We there use a special limit where domain walls reduce to gas of hard (soft)
 one-dimensional rods for the Abelian (non-Abelian) cases. In the simpler cases of the Abelian-Higgs model on a torus, our 
results agree with exact results which are geometrically derived by an explicit integration over the moduli space of 
the vortices.
On the other side of the limit, we can compute the volume of the moduli space in the combinatorial way, where the 
problem on the random (plane) partition appears as well as the four dimensional instanton calculus. 
A part of this talk is based on collaboration with
M.~Eto, T.~Fujimori, M.~Nitta, K.~Ohashi and N.~Sakai [hep-th/0703197].
} 
\maketitle
\section{Introduction}
\label{intro}

BPS solitons play essential roles
to understand the non-perturbative dynamics and properties
 in supersymmetric gauge theory.
It is an important task to evaluate the non-perturbative effects
from various kinds of BPS solitons, like instantons, monopoles, vortices and
domain-walls, which are classified by their codimensions.
To carry out this, we need to investigate the whole
detail structure (topology, metric, singularity, etc.)
of the moduli space of the solitons, but I would like to concentrate on calculations
of the ``volume'' of their moduli space in this talk.
In general, the moduli space of the solitons is non-compact and the volume of the moduli
space diverges. So we need a regularization in the calculation of the volume.
The volume in double-quotes means it is evaluated with a suitable regularization.

There are important and interesting applications  of the ``volume'' of the moduli
space of the BPS solitons. One of them appears in a thermodynamical partition
function of diluted BPS soliton gas. If we assume the BPS solitons do not interact
with each other and behave as free particles,  integration over phase space,
which is a cotangent bundle over the moduli space, reduces to integration over
the moduli space with suitable metric. Then the thermodynamical partition function
is proportional to the ``volume'' of the moduli space
\[
Z=\left(
\frac{T}{2\pi \hbar^2}
\right)^N
{\rm Vol}({\cal M}_N),
\]
where $T$ is temperature of the system and ${\rm Vol}({\cal M}_N)$
stands for the ``volume'' of the moduli space of $N$ BPS solitons.

Another application is proposed by Nekrasov \cite{Nekrasov:2002qd}.
He has shown that the prepotential of ${\cal N}{=}2$ supersymmetric gauge theory,
which includes all non-perturbative instanton corrections,
can be obtained from a statistical partition function summing over Young tableaux.
The partition function measures a regularized
``volume'' of the moduli space of $k$ instantons with gauge group $U(r)$
and the prepotential is given by a leading term of free energy
in an asymptotic expansion with respect to a regularization parameter, which
corresponds to the so-called $\Omega$-background.

In this talk, we propose a novel and simple derivation of the ``volume'' of the
moduli space of the BPS solitons, in particular BPS vortices, by evaluating or counting 
a pictorial configuration space of solitons.
The configuration space is realized by a D-brane configuration.
First, we consider the configuration space of vortices on a compact torus.
In this case, effective size of vortices affects the entire volume of the moduli space,
since the vortices can not overlap with each over due to their size.
We naively expect that the vortices behave like finite size disks
on the two-dimensional torus,
but the D-brane picture shows us that a configuration space of 
domain-walls, which is T-dual to the vortices, is equivalent to a configuration of
finite size hard rods on a one-dimensional circle in a special limit of parameters
\cite{Eto:2007aw}.
This limit is an approximation, but surprisingly the result is exact because of the
localization.

Secondary, we also consider a large area limit of the base space where the vortices exist.
In this limit, the effective size of the vortices can be ignored. So the configuration
space of the vortices or T-dual domain-walls reduces to a configuration of
rectangular kinks. We can identify the kink configurations with three-dimensional (3d)
Young diagrams (plane partitions).
So the evaluation of the volume of the moduli space reduces to
a counting of the skew Young tableaux associated with the kink configurations.
This is also an approximation, but we expect that this gives an exact answer. 
We can compare with a non-perturbative F-term of two-dimensional supersymmetric
gauge theory in evidence \cite{Shadchin:2006yz}.

\section{Thermodynamics of vortices on $T^2$}

We consider BPS vortices in supersymmetric $SU(N_c)$ gauge theory with 8 supercharges,
which has $N_f$ flavor hypermultiplets.
The theory possesses gauge fields $A_\mu$,  adjoint scalar fields $\Sigma$ in the vector multiplet and the hypermultiplets $H$ which
are expressed by an $N_c{\times}N_f$ matrix. After Bogomol'nyi completion of energy density,
we find BPS equation for the vortices
\[
{\cal D}_{\bar{z}} H = 0,\quad
F_{z\bar{z}} +\frac{g^2}{2}(c{\bf 1}_{N_c}-HH^\dag)=0,
\]
where $g$ is a gauge coupling constant and $c$ is a Fayet-Iliopoulos parameter.
If we consider solutions of the above BPS equations on a compact two-dimensional
surface, like a torus $T^2$, there exists a bound between the number of vortices $k$
and the area of the torus $A$
\[
k\frac{1}{N_c}\frac{4\pi}{g^2c} \leq A.
\]
This means that the vortex has a minimal effective area $\frac{1}{N_c}\frac{4\pi}{g^2c}$, which excludes the other vortices. This unit area is called the Bradlow area.

\begin{figure*}
\begin{center}
\includegraphics[width=0.8\textwidth,angle=0]{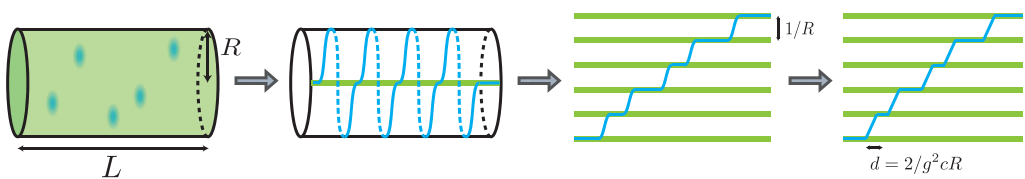}
\end{center}
\caption{Vortices as a bound state of D-branes. T-duality maps it to a kinky D-brane configuration. If we consider the configuration in a covering space, it is equivalent
to a system of hard rods in one-dimension.}
\label{T-dual}       
\end{figure*}

Now we realize these BPS vortices as D-brane bound states in Type IIA superstring
theory. For example, if we consider three-dimensional model, it is realized by a bound
state of $k$ D0-branes, $N_c$ D2-branes and $N_f$ D6-branes in 
$\R^{1,2}\times\C^2/\Z_2\times\R^3$, where the orbifold preserves a half of
 supercharges
and removes extra moduli (flat directions). As discussed in \cite{Eto:2006mz},
the T-duality maps a Wilson line of the gauge field which express the
vortex configuration to a domain-wall (kinky D-brane) configuration.
If we take a T-duality along one cycle (with radius $R$) of the torus,
 the $k$ vortices are mapped
to the kinky D-branes which wrap $k$-times on a dual cycle.
It is convenient to see the kinky D-brane configuration in covering space of the
dual cycle. In the covering space, the kinky D-branes is expressed as kink configurations
interpolating between the flavor D5-branes. In general, the kink configuration is smooth
functions, but if we take a special limit of $g^2c\to\infty$ and $1/R\to \infty$ with
$d\equiv 2/g^2c R$ fixed, then the shape of the kinks become sharp and is represented
by piece-wise linear functions.
 For example, if we consider the simplest
case, namely $k$-vortices with $N_c=N_f=1$ (ANO vortices), the above process is
depicted in Fig.\ref{T-dual}.

%

Once we take the limit and get the piece-wise linear configuration,
we can identify the domain-wall kink configuration with
a system of $k$ hard rods
with length $d$ on $S^1$ with radius $L$\footnote{
$L$ is a radius of another cycle of the torus against the T-dual direction.}.
Therefore, if we want to calculate the volume of the configuration (moduli)
space of the vortices, we need to calculate the volume of the T-dual
domain-wall
configuration space, which furthermore reduces to the volume of the configuration space
of hard rods on $S^1$. This reduced problem is easy to calculate. The answer to the
$k$ ANO vortices is proportional to $\frac{2\pi L(2 \pi L-kd)^{k-1}}{k!}$. The statistical 
partition function
of gas of $k$ identical hard rods with mass $m$ on $S^1$ with radius $L$
\[
Z_{\rm rods} = \left(
\frac{mT}{2\pi}
\right)^k
\frac{2\pi L(2\pi L-kd)^{k-1}}{k!}.
\]
Note that these parameters are expressed in terms of the dual picture. So,
in order to obtain the thermodynamical partition function of the original vortex system,
we replace $m$ with $(2\pi)^2 Rc$ and $d$ with $2/g^2cR$. Then we finally obtain
the partition function of the vortices
\be
Z_{k,T^2}^{N_c=N_f=1}=\frac{1}{k!}(cT)^k A\left(
A-k\frac{4\pi}{g^2c}
\right)^{k-1},
\label{partition function}
\ee
where $A=(2\pi)^2RL$ is the area of the torus.
From the partition function (\ref{partition function}), we can derive the van der
Waals equation of state of the vortex gas
\be
P\left(
A-k\frac{4\pi}{g^2c}
\right)=k T,
\label{EOS}
\ee
where
the Bradlow area $\frac{4\pi}{g^2c}$ appears and pressure diverges
at $A=k\frac{4\pi}{g^2c}$. This agrees with the arguments over the BPS equations.

In the above derivation, we have used the special limit of the parameters where
the kink configuration approximates to the piece-wise linear function. Thus
we can evaluate easily the volume of the moduli space. 
However, in spite of the approximate calculation,
 we find the results (\ref{partition function})  and (\ref{EOS}) are exact
as compared with \cite{Manton:2004tk}.
This means that a kind of localization works in the calculation of the volume of
the BPS vortex moduli space as like as in the instanton calculus, that is,
the volume does not depends on the detailed structure of the moduli space and
is determined by fixed point structure of isometries.
This localization reduces the problem to the simple statistical or combinatorial one.
We will see another evidence in the next section.

Our argument can be extended to the general $N_c$ and $N_f$ case.
The problem of the volume calculation of the moduli space also reduces
to the one-dimensional statistical system. For $N_f>1$, there appear short length
hard rods  which correspond to domain-walls connecting the Higgs vacua  
within a period of the covering space.  In the limit of $1/R\to \infty$, these shot rods are regarded as particles
bound on the long rods corresponding to the kinky D-branes wrapping around
the dual circle. In addition, the hard rods now can be overlapped with each other,
so the rods effectively become ``soft'' for the general non-Abelian case.  
Thus the configuration space of the non-Abelian vortices is equivalent to the
rods with the particles in one-dimension. It is however difficult to integrate
over the whole configuration space of the reduced system of the rods. 
We can perform it for the some special cases. For example, in the case of
$k$ local non-Abelian vortices with $\NC=\NF=N$, we obtain an expansion of the
partition function in terms of $1/A$
\[
{\small
\begin{split}
Z^{\NC=\NF=N}_{k,T^2} 
&=  \left( cT \right)^{kN} \frac{1}{k!} 
\left[\frac{A}{(N-1)!}\left(\frac{4\pi}{g^2c}\right)^{N-1}\right]^{k}\\
&\quad \times \left[ 1 - D_N (k-1) \frac{k}{A}  +
\mathcal O\left(\left(\frac{4\pi}{g^2c A}\right)^{2}\right) \right],
\end{split}
}
\]
where $\frac{D_N}{4\pi/g^2c} = \frac{(2N-2)!!}{(2N-1)!!}$.
The partition function for the 
$k$ semi-local vortices with $N_c=1$ and general $N_f$ is give by
{\small
\[
Z^{\NC=1,\NF}_{k,T^2} 
= \left( cT \right)^{k\NF} \frac{1}{k} \frac{1}{(k\NF-1) ! } 
A \left( A-\frac{4\pi k}{g^2c} \right)^{k\NF-1}.
\]}
We can also derive the equation of state for these vortex gases.

\section{Large area limit}

So far we have been treating the vortex system on the compact 2-dimensional
surface, but here we take a large area limit of the base space, namely
let us consider the vortices on $\C$. If we assume $k$ vortices behave
as point particles, the moduli space is a symmetric product space ${\cal M}_k
\simeq \C^k/S_k$,
where $S_k$ is the symmetric group of order $k$. The volume of the moduli space,
of course, diverges since $\C$ is non-compact, and is proportional to $({\rm Vol}(\C))^k$.
As in a spirit of \cite{Nekrasov:2002qd}, we expect that it makes a sense to pick up
a coefficient of the divergent volume, that is, a regularized ``volume'' has 
important information to investigate the structure of the vortex moduli space and
we can apply it to various physical problems.

\begin{figure}
\begin{center}
\includegraphics[width=0.22\textwidth,angle=0]{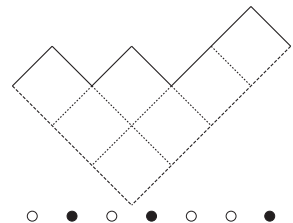}
\end{center}
\caption{D-brane bound states associated with the Higgs vacua relate to Young
diagrams via the Maya / Young diagram correspondence.
White and black circles represents D5-branes and D1-D5 bound states, respectively.}
\label{Higgs and Young}       
\end{figure}

Before explaining how to evaluate the regularized ``volume'',
we notice that there is one-to-one correspondence between
the Higgs vacua and the Young diagrams.
We are considering the supersymmetric $SU(N_c)$ Yang-Mills theory (SQCD)
with 8 supercharges. Assuming $N_c < N_f$ and $N_f$ matter hypermultiplets
have non-degenerate masses which are ordered as
$m_1<m_2<\cdots<m_{N_f}$.
The eigenvalues of the vev for the adjoint scalar in the Higgs phase
are given by choosing $N_c$ masses from $m_i$'s.
So the number of the Higgs vacua is ${}_{N_f}C_{N_c}=\frac{N_f!}{N_c!\tilde{N}_c!}$,
where $\tilde{N}_c=N_f-N_c$.
In the language of the brane configuration, the Higgs vacua correspond to
bound states of D1 and D5-branes, whose positions are related to
the adjoint scalar vev and hypermultiplet masses, respectively.
There works a kind of exclusion principle
and only one D1 can bind to D5 due to the orbifolding by $\Z_2$.
Identifying the D1 brane positions with right-up edges of the $-45^\circ$ rotated
Young diagram, that is, using the Maya diagram (free fermion Fock space) /
Young diagram correspondence,
the Higgs vacuum corresponds to a Young diagram within $N_c \times \tilde{N}_c$
boxes. (See Fig.\ref{Higgs and Young}.) Thus the Higgs vacuum can be
labeled by the Young diagram $\lambda$.

The domain-wall in this supersymmetric system interconnects two different Higgs vacua.
If we choose two different vacua (Young diagram) as $\lambda_1$ and $\lambda_2$
and determine orientation of a ``time'' direction which is transverse to the
domain-wall, the BPS condition says that the Young diagram $\lambda_1$
should be include inside the $\lambda_2$, namely $\lambda_2$ must be constructed
by just adding some boxes to $\lambda_1$. We represent this inclusion relation
by $\lambda_1 \prec \lambda_2$. If we arrange $k$ domain-walls
by a ``time'' series, the $k+1$ vacua $(\lambda_1,\lambda_2,\ldots,\lambda_{k+1})$
should satisfy
\[
\lambda_1 \prec \lambda_2 \prec \cdots \prec \lambda_{k+1}.
\]
This means that $k$ BPS domain-walls consists a plane partition (3d
Young diagram). This correspondence, however, is abstract since
the domain-walls have thickness and transition between vacua is smooth.
If we take a limit of $g^2c\to \infty$, the transition becomes steep
since the kink profile of the domain-wall gets rectangular (similar
to the limit of $d\to 0$ in the previous section).
Thus the correspondence between the BPS domain-walls and 3d Young diagram
is exact in the above limit. We will consider this situation in the following.

\begin{figure}
\begin{center}
\includegraphics[width=0.40\textwidth,angle=0]{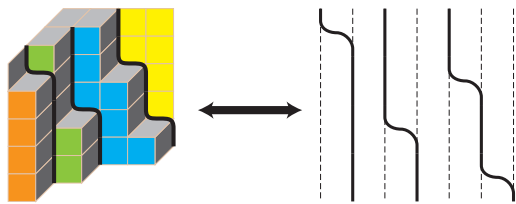}
\end{center}
\caption{There is one-to-one correspondence between 3d Young diagrams
and kink flows (kinky D-branes).}
\label{Kink and 3d Young}       
\end{figure}

Heights of the 3d Young diagram relate to the positions of the domain-walls,
or we can equivalently express the positions as numerical numbers inside
the boxes of a skew Young diagram $\lambda_{k+1}/\lambda_1$,
which is removal of boxes of $\lambda_1$ from ones of $\lambda_{k+1}$.
The number of domain-walls is given by $k=|\lambda_{k+1}|-|\lambda_1|$, where $|\lambda|$ stands for the number of the boxes in the Young diagram $\lambda$.
We call the skew Young diagram with the numerical numbers inside the boxes
the skew Young tableau.
In order to calculate the volume of the moduli (configuration) space of the
domain-walls, we need to count the number of all possibilities of the Young
tableau. It is similar to evaluation of a dimension of a symmetric group
associated with the skew Young diagram, but it diverges in our case
since the transverse direction to the domain-walls is continuous and non-compact.
To regularize the counting of the domain-wall configuration space, we
consider a finite interval, discretize the transverse direction,
and label the positions by integers from one to sufficiently large $N$.
Therefore, we find the volume of the domain-wall moduli space
${\cal M}^{\rm DW}_k$
 is given by
{\small\[
{\rm Vol}({\cal M}^{\rm DW}_k)
=\lim_{N\to \infty, q\to 1}\frac{1}{N^k}
s_{\lambda_{k+1}/\lambda_1}(1,q,q^2,\ldots,q^{N-1}),
\]}
where $s_{\lambda/\lambda'}(x_1,x_2,\ldots,x_N)$ is the skew Schur function.
If we simply choose $\lambda_1=\emptyset$ and $\lambda_{k+1}=\lambda$,
then
\[
\begin{split}
{\rm Vol}({\cal M}^{\rm DW}_k)
&=\frac{d_\lambda}{k!}
=\prod_{i<j}
\frac{\mu_i-\mu_j-i+j}{-i+j}
,
\end{split}
\]
where $d_\lambda$ is the dimension of the symmetric group and
$\mu_i$,
 which satisfies $\mu_1 \geq \mu_2 \geq \cdots$ and $\sum_{i=1}^\infty \mu_i =k$, 
is the number of the boxes in the $i$-th row of the Young diagram $\lambda$.
This volume is related to the volume
of the moduli space of the large $N$ $U(N)$ flat connections on the two-dimensional disk \cite{Matsuo:2004cq}.
It also corresponds to just a ``half'' of the moduli space volume of
non-commutative $U(1)$ instantons. 
This fact is reminiscent of the observation by Hanany and Tong \cite{Hanany:2003hp}.

To apply the above domain-wall result to the vortex, we have to consider the
multiple domain-wall configuration in the dual covering space as similar to
the previous section. For example, the $k$ ANO vortices in the
$N_c=N_f=1$ theory are equivalent to $k$ domain-walls of the $N_c=1$ and $N_f\to\infty$ by T-duality. Indeed, if we count the number of configurations of $k$ discrete
positions in $N$, it gives
\[
{\rm Vol}({{\cal M}^{\rm ANO}_{k,\C}})=\lim_{N\to\infty}\frac{1}{N^k}\frac{N!}{k!(N-k)!}
=\frac{1}{k!}.
\]
This agrees with the large area limit of (\ref{partition function})
by identifying the divergent area $A^k$ with $N^k$.
Similarly we can count the dual domain-wall configurations to the vortices
for the particular cases in the combinatorial way. It precisely gives
a coefficient of the large area divergent part.

\section{Conclusion and Discussions}

In this talk, I give a simple schematic and combinatorial calculation of the
volume of the vortex and domain-wall moduli space.
These BPS objects have codimension 2 and 1.
The problem of the moduli space volume calculation reduces to
the evaluation of the volume of the one-dimensional hard rod configuration space,
or the counting of the number of 3d Young diagrams (plane partitions).
These calculations also related to the counting of the BPS D-brane configurations
in the realization of supersymmetric gauge theory. We expect that these
schematic and D-brane picture can be applied to other solitons like
instantons and monopoles. 

On the other hand, the calculation of the moduli space volume has another application
to the derivation of the effective prepotential or superpotential (F-terms)
in supersymmetric gauge theory. 
For two-dimensional supersymmetric Yang-Mills theory, the non-perturbative
effective superpotential is calculated in \cite{Shadchin:2006yz} by using equivariant cohomology
or topological matrix model. This is an effective theory in the Coulomb phase,
but similar combinatorial objects (Gamma functions) appears in the partition function,
even though our calculation is done in the Higgs phase.
We also hope that our calculation sheds light on the relation between effective
theories in the Coulomb and Higgs phase, and makes clear non-perturbative dynamics
in supersymmetric gauge theories.

%
%

\end{document}